\documentclass[aps,prd,onecolumn,11pt,notitlepage,nofootinbib,superscriptaddress,showkeys,letterpaper]{revtex4-1}
\usepackage{amstext,amsmath,amssymb,amsfonts,bbm, amsthm}

\usepackage[latin1]{inputenc}

\usepackage{fancyhdr}
\usepackage{graphicx}
\usepackage{hyperref}

\usepackage{epstopdf}
\usepackage{rotating}
\usepackage{cleveref}

\usepackage{float}

\usepackage{cases}
\usepackage{tabls}

\usepackage{subfigure}

\usepackage{amsthm}

\newtheorem{theorem}{Theorem}

\usepackage{tocvsec2}

\usepackage{color}

\def\beq{\begin{equation}}
\def\be{\begin{equation}}
\def\ee{\end{equation}}
\def\bes{\begin{eqnarray}}
\def\ees{\end{eqnarray}}

\begin{document}
%%%%%%%%%%%%%%%%%%%%%%%%%%%%%%%%%%%%%%%%%%%%%%%%%%%

\title{Natural selection as coarsening}%: Evolutionary Biology Meets Material Science}

\author{Matteo Smerlak}\email{msmerlak@perimeterinstitute.ca}
\affiliation{Perimeter Institute for Theoretical Physics, 31 Caroline St.~N., Waterloo ON N2L 2Y5, Canada}

\date{\small\today}
%----------------------------------------------------------------------------------------
%	ABSTRACT, KEYWORDS AND ABBREVIATIONS
%----------------------------------------------------------------------------------------

\begin{abstract}

Analogies between evolutionary dynamics and statistical mechanics, such as Fisher's  second-law-like "fundamental theorem of natural selection" and Wright's ``fitness landscapes", have had a deep and fruitful influence on the development of evolutionary theory. Here I discuss a new conceptual link between evolution and statistical physics. I argue that natural selection can be viewed as a \textit{coarsening phenomenon}, similar to the growth of domain size in quenched magnets or to Ostwald ripening in alloys and emulsions. In particular, I show that the most remarkable features of coarsening---scaling and self-similarity---have strict equivalents in evolutionary dynamics. This analogy has three main virtues: it brings a set of well-developed mathematical tools to bear on evolutionary dynamics; it suggests new problems in theoretical evolution; and it provides coarsening physics with a new exactly soluble model. 
\end{abstract}

\maketitle
%------------------------------------------------

\section{Introduction}
Like statistical mechanics, evolutionary theory deals with the macroscopic transformations and statistical regularities of large sets of individual units (molecules and replicators, respectively). It is therefore unsurprising that conceptual and formal analogies can be drawn between these two fields; indeed, many authors have pursued these links, see e.g. Refs. \cite{Iwasa:1988jza,Franz:1993ep,Barton:2009di,Mustonen:2010igb}. These works are often inspired by two ideas introduced by the co-founders of population genetics: the  ``fundamental theorem of natural selection" derived by Fisher \cite{Fisher:1930tj}, who viewed mean fitness as an entropy-like function for evolution, and Wright's ``adaptive landscape", which is reminiscent of the physicist's potential landscape with fitness playing the role of minus energy. These ideas---Fisher's \textit{fitness as entropy} and Wright's \textit{fitness as (minus) energy}---have been reviewed in many places; I refer the reader to the literature for details on the history, conceptual clarity and heuristic value of these two seminal analogies.\footnote{In particular, we refrain from dwelling on the seemingly contradictory nature of Fisher's and Wright's metaphors, at least in the physicist's eye: how can fitness be both like entropy and like energy?}

My purpose in this paper is to discuss a third and distinct analogy between evolution and physics. In short, I propose to view \textit{natural selection as coarsening}. Natural selection is the principle of dominance of the best replicators and is the cornerstone of Darwin's theory of evolution; coarsening is the growth of large structures in heterogenous mixtures, solid solutions, emulsions, etc. As I will now show, these two concepts are closely related, both conceptually and formally.

\section{A brief reminder on coarsening phenomena}

Coarsening refers to any relaxation process wherein the characteristic length scale grows over time \cite{Krapivsky:2010dg}. Examples of coarsening phenomena abound in condensed matter physics: domain growth in quenched magnets;  Ostwald ripening in alloys and emulsions; bubble coalescence in soap froths; phase separation in binary mixtures;  etc. In these systems, excess free energy is stored in localized defects (like domain walls or bubble interfaces) whose size spontaneously increases over time so as to reduce their density and hence the total free energy. The relevance of coarsening phenomena extends to astrophysics (galactic clustering), socio-dynamics (consensus formation, racial segregation), and in many other branches of sciences. And not just science: Ostwald ripening is the reason why old ice creams taste gritty and pastis and ouzo look cloudy. 

Key universal features of coarsening are \textit{dynamic scaling} and \textit{asymptotic self-similarity} \cite{Krapivsky:2010dg}. Dynamic scaling refers to the power-law growth $L(t)\sim t^{1/z}$ of the characteristic length scale $L(t)$, where $z$ is a dynamical scaling exponent which depends on spatial dimension, conservation laws, etc., but not on the specifics of the system under study. (For ``conserved order parameters" we typically have $z=3$; for ``non-conserved order parameters", we usually find instead $z=2$.)  Asymptotic self-similarity means that the distribution of sizes in the system $\ell$ becomes invariant under the space- and time-rescaling $\ell\mapsto\lambda^{1/z}\ell$ and $t\mapsto \lambda t$. The emergence of scaling and self-similarity reflects the ``endogenous" nature of coarsening: rather than being driven externally to match any given, fixed external scale, spatial structures evolve to correct initial heterogeneities among themselves. No scale is intrinsically preferred in this problem. 

One of the simplest (and oldest) formulation of coarsening dynamics is the Lifshitz-Slyozov-Wagner (LSW) mean-field model \cite{Lifshitz:1961cc,Wagner:1961gz} of Ostwald ripening. This model describes the evolution of an ensemble of spherical particle clusters through the dissolution and redeposition of small clusters onto larger ones, at fixed total cluster volume. The concentration of clusters with volume $v>0$ at time $t\geq0$ satisfies the (non-local) continuity equation\footnote{In suitable units, see the references cited above for details.}
\begin{equation}\label{LSW}
	\frac{\partial c_t(v)}{\partial t}=\frac{\partial}{\partial v}\left[\left(\frac{v^{1/3}}{L(t)}-1\right)c_t(v)\right],
\end{equation} 
where the length scale $L(t)$ is the mean cluster size, defined as
\begin{equation}
	L(t)\equiv \int_0^\infty v^{1/3}c_t(v)dv\Bigg/\int_0^\infty c_t(v)dv.
\end{equation}  
LSW showed from these equations that $L(t)\sim t^{1/3}$ and that generic solutions $c_t(v)$ can be rescaled (by the mean particle size) to approach a fixed limiting size distribution $c^*_\infty(v)$. In fact, as discussed \textit{e.g.} in \cite{Giron:1998cn}, the LSW scaling solution $c^*_\infty(v)$ is only one (extremal) member of a one-parameter family of limiting size distributions $c^*_\theta(v)$; the basin of attraction of each $c^*_\theta(v)$, as well as the multiplicative constant in the scaling relation $L(t)\sim t^{1/3}$, is determined by the large-size behavior of the initial distribution \cite{Niethammer:1999hg}. In essence, $c_t(v)$ can be rescaled to converge to $c^*_\theta(v)$ if \cite{Niethammer:1999hg}%\footnote{The necessary and sufficient condition is that the survival function on the left-hand side is regularly varying with index $p$ when $v\to v_\textrm{m}$. Regular variation is a central concept in analysis and probability theory.}
\begin{equation}
	\int_v^{v_\textrm{m}} c_0(u)du\sim (v_\textrm{m}-v)^\theta \quad\quad \textrm{when}\ v\to v_\textrm{m},
\end{equation}
where $v_\textrm{m}$ is the volume of the largest cluster at $t=0$. The original LSW solution corresponds to $\theta\to\infty$. The interpretation of these results is intuitive: at late times, all clusters but the largest ones have disappeared. For this reason the asymptotic size distribution $c_t(v)$ is completely characterized by the large-size \textit{tail} of the initial distribution $c_0(v)$. The (conserved) tail exponent $\theta$ captures this tail behavior. 

\section{Dynamic scaling and self-similarity in selection dynamics}

Let us now introduce the basic concept underlying natural selection---\textit{fitness}. Fitness is usually defined either as the rate of exponential growth of lineages\footnote{Lineage: the descendants of a common ancestor.} (Malthusian fitness) or as the expected number of offspring per replicator\footnote{Replicator: anything (organisms, genes, etc.) that replicates itself.} per generation (Wrightian fitness). Moreover, fitness is typically discussed as a function of a genotype, or of a genotype plus an environmental configuration. These definitions make sense from a biological perspective, but they are imperfect from a conceptual standpoint, because neither growth nor replication nor heredity are required for natural selection to act. All that is required is that a population can be divided into distinct types $\tau$, such that the numbers of individuals of any two types $\tau_1$ and $\tau_2$ over time $t$ satisfy
\begin{equation}\label{deffitness}
	\frac{N_t(\tau_1)}{N_t(\tau_2)}=\exp\left[\big(x(\tau_1)-x(\tau_2)\big)t\right]
\end{equation}  
for some function (defined up to a constant) $x(\tau)$. This function is fitness, and the \textit{relative} growth of high-fitness types---the ``dominance of the fittest"---is natural selection. Whether the total population size actually grows, remain constant or shrinks; whether individuals replicate or simply die at different rates; and whether they carry a genotype or not, is irrelevant for the dynamics of natural selection. The only fundamental structure in natural selection is the distribution of fitness $p_t(x)$, and its mathematical study boils down to the analysis of the (infinite dimensional) flow $p_t(x)$. Once we acknowledge this basic fact, the analogy with coarsening becomes immediately apparent: in coarsening, the number of large clusters grows at the expense of smaller ones; in natural selection, the number of high-fitness individuals grows at the expense of lower-fitness ones. Put succintly, natural selection is coarsening in  fitness space.  

What equation does $p_t(x)$ satisfy? From \eqref{deffitness}, the number $N_t(x)$ of individuals with fitness $x$ evolves as
\begin{equation}
	N_t(x)=M_t\, e^{xt}
\end{equation}
for some function of time $M_t$. As a result, the distribution of fitness $p_t(x)=N_t(x)/\int N_t(y)dy$ satisfies
\begin{equation}\label{tilt}
	p_t(x)=\frac{p_0(x)\,e^{xt}}{Z_t},
\end{equation}
where $Z_t=\int_{-\infty}^{\infty} p_0(y)e^{yt}dy$. Taking time derivatives, this corresponds to the non-linear integro-differential equation 
\begin{equation}\label{selection}
	\frac{\partial p_t(x)}{\partial t}=(x-\mu_t)p_t(x),
\end{equation}
in which 
\begin{equation}\label{meanfitness}
	\mu_t\equiv \int_{-\infty}^{\infty}yp_t(y)dy
\end{equation}
is the mean fitness at time $t$. Eq. \eqref{selection} is the fundamental equation of natural selection. 

In the past, the selection equation \eqref{selection} has often been analyzed from the perspective of the \textit{cumulants} of $p_t(x)$, which are the derivatives at $\omega=0$ of the generating function
\begin{equation}
	\psi_t(\omega)\equiv \int e^{\omega x}p_t(x)dx.
\end{equation}
Indeed, it is easy to see that, if $\kappa_t^{(m)}$ denotes the $m$-th cumulant of $p_t(x)$, \eqref{selection} is equivalent to the tower of equations 
\begin{equation}\label{cumulanttower}
	\frac{d\kappa_t^{(m)}}{dt}=\kappa_t^{(m+1)} \quad \textrm{for}\ m\in \mathbb{N}. 	
\end{equation}
In particular, the time-derivative of the mean fitness $\mu_t=\kappa_t^{(1)}$ is equal to the variance in fitness $\kappa_t^{(2)}$, which is the---much commented \cite{Ewens:1989hca,EDWARDS:1994gd,Frank:1997fb}---Fisher ``fundamental theorem of natural selection" \cite{Fisher:1930tj}. 

The problem with this approach, of course, is that the infinite tower \eqref{cumulanttower} does not close. This has led to some debate about the true meaning of Fisher's theorem, its ``dynamic sufficiency" \cite{Lewontin:1974te,Barton:1987hla}, etc. Instead of truncating \eqref{cumulanttower} at some finite $m$ \cite{Gerrish:2011wz}, a fruitful approach is to deal with the generating function itself, noting with Eshel \cite{Eshel:1971ura} that
\begin{equation}
	\psi_t(\omega)=\psi_0(\omega+t)-\psi_0(t).
\end{equation}
From this observation it follows that the late-time dynamics of natural selection reduces to the asymptotic behavior of the initial generating function $\psi_0(\omega)$. Using general results in asymptotic analysis,  Youssef and I obtained the following results \cite{Smerlak:2017ez,Smerlak:2017vf}:\footnote{We later learnt that (parts of) this result were already known in the statistical literature \cite{Balkema:2003hnb}, because families of distributions of the form \eqref{tilt} (called exponential families) play an important role there.}

\begin{theorem}
Let $x_\textrm{m}$ denote the upper endpoint of the support of $p_0(x)$, $F_0(x)\equiv \int_x^{x_\textrm{m}}p_0(y)dy$ its survival function, $\mu_t$ (resp. $\sigma_t$) the mean (resp. standard deviation) of $p_t(x)$ and $\overline{p}_t(x)\equiv \sigma_tp_t(\sigma_t x+\mu_t)$ the standardized fitness distribution. The late-time behavior of the selection equation \eqref{selection} satisfies
\begin{itemize}
	\item If $x_\textrm{m}=+\infty$ and $-\ln F_0(x)\underset{x\to\infty}{\sim} Ae^{\alpha x}$ for some $\alpha>0$, then $\mu_t\sim \ln t/\alpha$, $\sigma_t^2\sim 1/\alpha t$ and $\overline{p}_t(x)$ converges to the standard Gaussian as $t\to\infty$. 
	\item If $x_\textrm{m}=+\infty$ and $-\ln F_0(x)\underset{x\to\infty}{\sim} Bx^{\beta}$ for some $\beta>0$, then $\mu_t\sim B'\beta' t^{\beta'-1}$, $\sigma_t^2\sim B'\beta'(\beta'-1)\,t^{\beta'-2}$ and $\overline{p}_t(x)$ converges to the standard Gaussian as $t\to\infty$. Here $B'\equiv(\beta-1)(\beta B)^{-1/(\beta-1)}/\beta$ and $\beta'\equiv\beta/(\beta-1)$.
	\item If $x_\textrm{m}<\infty$ and $F_0(x)\underset{x\to x_\textrm{m}}{\sim} C(x_\textrm{m}-x)^\gamma$ for some $\gamma>0$, then $\mu_t\sim \ln w_+-\gamma/t$, $\sigma_t^2\sim \gamma/t^2$, and $\overline{p}_t(x)$ converges to the flipped gamma distribution with density $$p^*_\gamma(x)\equiv 		\frac{\gamma^{\gamma/2}}{\Gamma(\gamma)}\,e^{-\sqrt{\gamma}(\sqrt{\gamma}-x)}(\sqrt{\gamma}-x)^\gamma$$
and support $(-\infty,\sqrt{\gamma}]$ as $t\to\infty$. (The limiting distributions $p^*_\gamma(x)$ approach the Gaussian when $\gamma\to\infty$.)
\end{itemize}
\end{theorem}

These results establish asymptotic self-similarity for natural selection dynamics. They are clearly reminiscent of the behavior of the LSW equation; in particular, both admit a continuous one-parameter of (standardized) limiting distribution whose shape is determined by the tail thickness of the initial distribution. There are also differences with the LSW equation. Contrary to the latter, the growth exponent (and not just the prefactor) can depend---albeit weakly, via the tail structure only---on the initial condition. From that perspective, natural selection reveals subtleties in coarsening dynamics that are not present in the more classical models. 

The study of coarsening dynamics is generally complicated by the non-linear nature of the underlying dynamical equation. The LSW equation, for instance, is often considered one of the simplest models of coarsening. But the LSW equation---a non-local transport equation---is not particularly easy to handle mathematically: even proving existence and uniqueness of solutions is non-trivial \cite{Collet:2000bi,Niethammer:2000dm}. This is in contrast with the natural selection equation, which is (trivially) exactly soluble, see \eqref{tilt}. It could therefore be useful to think of natural selection as a particularly simple model of coarsening, indeed much simpler than the LSW equation; it goes without saying that the scientific importance of natural selection also compares favorably with that of Ostwald ripening. 
 
\section{Natural selection in a broader landscape}

Analogies are only as good as the new avenues they open for research. In this section I discuss two ways in which natural selection and coarsening can be viewed as part of a broader conceptual landscape: from the perspective of \textit{extreme value}, and from the perspective of \textit{dissipation}. It is possible that both fields could benefit from these broader perspectives.

\subsection{Extreme values and regular variation}

From a mathematical perspective, coarsening and natural selection are both concerned with probability distributions with ``growing tails": the features of the evolved (size or fitness) distributions become more and more dominated by the (large size or high fitness) tail of the initial data. Such tail-driven flows are found in other branches of probability theory, two classical examples being the generalized central limit theorem (CLT) \cite{Bouchaud:1990be} and Fisher-Tippett-Gnedenko (FTG) theorem of extreme value theory \cite{deHaan:2007bza}.\footnote{The conceptual link between coarsening dynamics and the central limit theorem was emphasized by Pego in his lecture notes \cite{Pego:2007hl}.}

The generalized CLT deals with sums of i.i.d. random variables $X_i$ that are \textit{fat-tailed}. Unlike the more familiar CLT, in which the $X_i$ are assumed to have a finite variance and the sum $S_n\equiv\sum_{i=1}^n X_i$ is dominated by typical values of $X_i$, the generalized CLT shows that, when $X_i$ is fat-tailed, the sum $S_n$ is increasingly dominated by rare, large fluctuations of $X_i$ when $n\to\infty$. In the limit, the scaling behavior and limiting distribution of $S_n$ are completely determined by a single number measuring the tail thickness of $X_i$. The same patterns holds in extreme value theory, where one studies the behavior of $M_n\equiv\max\,\{X_i\}_{i=1}^n$ for i.i.d variables $X_i$. According to the FTG theorem, the scaling behavior and limiting distribution of $M_n$ are again determined by a single tail-thickness index---the same index as in the generalized CLT.

How is this tail index defined? The answer lies within the notion of \textit{regular variation}. A function $f(x)$ is said to be regularly varying at infinity with index $\theta$ if it can be written 
\begin{equation}
	f(x)=x^\theta L(x)
\end{equation}
where $L(x)$ is such that $\lim_{x\to\infty}L(\lambda x)/L(x)=1$ for any $\lambda>0$. This definition can be extended to finite endpoints: we say that $f(x)$ is regularly varying at $x=x_\textrm{m}<\infty$ with index $\theta$ if $f(x_\textrm{m}-1/x)$ is regularly varying at infinity with index $\theta$. Regular variation is just the right concept for both the generalized CLT and the FTG theorem: necessary and sufficient conditions for the convergence of both $S_n$ and $M_n$ to their respective limiting distributions involve regular variation; the ``tail index" evoked above is the index of regular variation of the distribution function of $X_i$.  

This is also true in coarsening and natural selection dynamics. In the context of the LSW equation, Niethammer and Pego have argued that $\overline{c}_t(x)$ converges to a limiting type $c^*_\theta(x)$ iff the survival function of $c_0(x)$ is regularly varying at its upper endpoint with index $\theta$ \cite{Niethammer:1999hg}; see also \cite{Menon:2004fa} for a comprehensive analogy between coarsening and the generalized CLT (this time in the context of the Smoluchowski equation, a relative of the LSW equation). For natural selection (in the case $x_\textrm{m}<\infty$), the necessary and sufficient on $F_0(x)$ for $\overline{p}_t(x)$ to converge to a flipped gamma distribution is that it be regularly varying at $x_\textrm{m}$ \cite{Balkema:2003hnb}. These common structures underscores the mathematical unity of coarsening dynamics. 

More broadly, the generalized CLT, the FTG theorem, the Niethammer-Pego result on LSW coarsening and the above results on natural selection should all been seen as variations on the theme of extreme values; I think of them as part of a broadly construed \textit{extreme value theory}\footnote{The phrase ``extreme value theory" is usually used in the restricted context of max-stable distributions.}  unified by the analytical concept of regular variation.

\subsection{Dissipation, Lyapunov functions and gradient flows}

Dissipation processes erase information over time. While this is usually meant in a physical sense, e.g. as the generation of heat in mechanical devices, nothing prevents us from interpreting the term in a more general sense. Coarsening, for instance, can be described as the erasure of the information in the distribution of cluster sizes through the growth of the largest clusters; similarly, natural selection can be described as the erase of the information encoded in low-fitness individuals. Mathematically, this corresponds to the existence of entire basins of attraction that all flow to the same late-time attractors, the limiting size and fitness distributions. But what kind of dissipative process are coarsening and natural selection?

A fruitful perspective on dissipative dynamics can sometimes be gained by interpreting the relevant dynamical equations as \textit{gradient flows} \cite{Ambrosio:2008co}. This means that the latter can be written as 
\begin{equation}
	\frac{\partial p_t(x)}{\partial t}=\nabla \mathcal{F}[p_t]
\end{equation} 
where $\mathcal{F}$ is a functional on the space of probability distributions and $\nabla$ is the gradient with respect to a suitable Riemannian structure on that space. This approach was pioneered in the context of Fokker-Planck equation by Otto and his collaborators \cite{Jordan:1997el}, who linked dissipation with the Wasserstein geometry of optimal transportation \cite{Villani:2003th}. For the simplest case of pure diffusion (i.e. $\partial p_t(x)/\partial t=\Delta p_t$), the functional $\mathcal{F}$ is nothing but the entropy $S[p_t]$ of $p_t$. One of the several advantages of such a reformulation is that it provides a Lyapunov functional (namely $\mathcal{F}$ itself) for the flow. If Boltzmann had not already done that a century and a half earlier from physical arguments, Otto \textit{et al.} might have discovered the link between dissipation and entropy growth through Wasserstein geometry!

Such geometric structures show their full potential when they also explain the emergence of universal limiting distributions. In the case of simple diffusion the rescaled density $\overline{p_t}(x)\equiv t^{1/2}p_t(t^{1/2}x)$ converges to a Gaussian distribution when $t\to \infty$, a simple application of the central limit theorem. Otto's Wassertein-geometric perspective illuminates this behavior: the dynamical equation for $\overline{p}_t(x)$ is the Wasserstein gradient flow for $S+V/2$, where $V$ denotes  variance. Because the variance of $\overline{p}_t(x)$ is fixed by construction, this implies that $S[\overline{p}_t]$ is monotonically increasing with $t$---hence $\overline{p}_t(x)$ must converge to the maximum-entropy distribution at fixed variance, the Gaussian. This entropic interpretation of the central limit theorem generalizes to any sum of i.i.d. variables (with finite entropy), a result proved in 2004 by Arstein \textit{et al.} \cite{Artstein:2004wj}. 

In the context of natural selection, it is known \cite{Hofbauer:1998wn} that \eqref{selection} is a gradient flow for the Fisher metric on the space of probability distributions
\begin{equation}
	\langle u,v\rangle_p \equiv \int \frac{u(x)v(x)}{p(x)}dx
\end{equation}
where $u(x)$ and $v(x)$ are two `tangent functions' (with vanishing integral). In this formulation, the function $\mathcal{F}[p_t]$ is nothing but the mean fitness $\mu_t$. The discussion above then suggests the following questions:

\begin{quote}
\begin{itemize}
	\item \textit{Can the evolution of standardized fitness distributions $\overline{p}_t(x)$ also be viewed as a gradient flow?}
	\item \textit{More generally, is the convergence of standardized fitness distributions to their limiting distribution monotonic? If so, in what sense?}
\end{itemize}
\end{quote}
Answering these questions would amount to \textit{explaining}, as opposed to simply \textit{proving}, the emergence of universality in natural selection, in the same way the $H$-theorem explains the universality of Boltzmann distributions at thermal equilibrium. The same questions can of course be asked for other models of coarsening. (For instance, a gradient flow structure for the LSW equation was identified by Niethammer in \cite{Niethammer:2004un}.)

\section{Conclusion}

Analogies are the soul of statistical mechanics. In the context of evolutionary dynamics, much of the quantitative work has been strongly inspired by two famous analogies: Fisher's tentative link between natural selection and the second law of thermodynamics, and Wright's use of adaptive landscapes.

In this paper, I have sketched another, potentially useful analogy between evolution and physics. I have linked natural selection with the dynamics of coarsening familiar from material science and mediterranean aperitifs.  Taking an even broader perspective, I have argued that both processes can be seen as instances of a generalized ``extreme values theory", and I have speculated that geometric concepts developed in the context of dissipative physics can perhaps be brought to bear to these problems.    

\section*{Acknowledgments}
The analogy between selection and Ostwald ripening was pointed out to me by Felix Otto during a visit at the Max Planck Institute for Mathematics in the Sciences. I thank Robert Pego and Baruch Meerson for comments on this manuscript. Research at the Perimeter Institute is supported in part by the Government of Canada through Industry Canada and by the Province of Ontario through the Ministry of Research and Innovation.

%\bibliographystyle{utcaps}
%\bibliography{library}

\providecommand{\href}[2]{#2}\begingroup\raggedright\endgroup

\end{document}